\documentclass[amssymb,12pt,longbibliography,nofootinbib]{revtex4-1}
\usepackage{graphicx}
\usepackage{appendix} 
\usepackage{hyperref}
\usepackage{bm,amsmath}
\usepackage{amsthm}
\usepackage{setspace}
\usepackage{slashed,mathrsfs}
\usepackage{xcolor}

\def\journal{Sci.~Rep.~{\bf 13}, 14106 (2023)}
\def\link{https://www.nature.com/articles/s41598-023-40893-0}

\def\BLO{ B ({\cal H}, {\mathbb C}) }

\def\PSD{ {\rm Her}^{\ge 0} ({\cal H}, {\mathbb C}) }

\newcommand{\KET}[1] {\left| #1 \right\rangle}
\begin{document}
\rightline{\href{\link}{\tt \journal}}
\vspace{0.2in}
\title{Proposal for a Lorenz qubit}
\author{\textsc{Michael R. Geller}}
\affiliation{Center for Simulational Physics, University of Georgia, Athens, Georgia 30602, USA}
\date{September 28, 2023}
\begin{abstract}
\vskip 0.1in
\begin{spacing}{0.9}
\centerline{\large \bf Abstract}
\vskip 0.3in
Nonlinear qubit master equations have recently been shown to exhibit rich dynamical phenomena such as period doubling, Hopf bifurcation, and strange attractors usually associated with classical nonlinear systems. Here we investigate nonlinear qubit models that support tunable Lorenz attractors. A Lorenz qubit could be realized experimentally by combining qubit torsion, generated by real or simulated mean field dynamics, with linear amplification and dissipation. This would extend engineered Lorenz systems to the quantum regime, allowing for their direct experimental study and possible application to quantum information processing.
\end{spacing}
\end{abstract}
\maketitle

Several recent papers \cite{KowalskAP19,190709349,200309170,KowalskiQIP20,210308982,211105977,211113477}
have considered nonlinear generalizations of the Gorini-Kossakowski-Sudarshan-Lindblad (GKSL) master equation \cite{GoriniJMP76,LindbladCMP76} for qudits. The superoperators resulting from these evolutions each take the form of a positive trace-preserving (PTP) channel \cite{SudarshanPR61,KrausAnnPhys71} $X \mapsto \phi(X)/{\rm tr}[\phi(X)]$, with $X$ a density matrix and $\phi$ a positive map.\footnote{Specifically, $\phi : \BLO \rightarrow  \BLO $ is a linear or nonlinear map on bounded linear operators satisfying: (i) $\phi(X)^\dagger = \phi(X^\dagger)$ for every $X \in \BLO$; 
(ii) $\phi(X) \succeq 0 $  for every $X \in \PSD$;
(iii) ${\rm tr}[\phi(X)] \neq 0$ for every $X \in \PSD$.
Here $\BLO$ is the complex vector space of bounded linear operators  $X : {\cal H} \rightarrow {\cal H} $ on our Hilbert space ${\cal H}$, and $\PSD = \{ X \in \BLO : X = X^\dagger, X \succeq 0 \} $ is the positive semidefinite (PSD) subset.}
The positivity of this {\it normalized} PTP channel follows from the positivity of $\phi$ and ${\rm tr}[\phi(X)] > 0$. It's trace preservation property is actually a trace fixing one, but these are physically equivalent when applied to normalized initial states. Kowalski and Rembieli\'nski \cite{KowalskAP19}, and also Rembieli\'nski and Caban \cite{200309170}, considered cases with linear $\phi$ and ${\rm tr}[\phi(X)] \neq 1$, extending Gisin's 1981 model \cite{GisinJPA81} to mixed states. We call these channels  {\it nonlinear in normalization only} (NINO) to emphasize that the nonlinearity in this case serves only to conserve trace. We might think of NINO channels as being ``mildly'' nonlinear. In particular, they satisfy a convex quasilinearity property \cite{190603869}, preventing superluminal signaling \cite{GisinPLA90,PolchinskiPRL91,CzachorFPL91,GisinJPA95,KentPRA05}.
The main difference between linear CPTP and NINO channels are that the generators of linear CPTP evolution are negative definite, leading to strictly nonexpansive dynamics, whereas NINO channels support non-CP \cite{PechukasPRL94,ShajiPLA05,CarteretPRA08,13120908} and entropy decreasing \cite{KowalskAP19} processes that amplify the Bloch vector \cite{KowalskAP19,211105977}. Hence we can interpret the NINO master equation as extending the GKSL equation to non-Hermitian Hamiltonians. Fernengel and Drossel \cite{190709349} studied cases where $\phi$ is nonlinear and ${\rm tr}[\phi(X)]=1$, a family of state-dependent CPTP channels obtained by adding state-dependence to a Hamiltonian and set of Lindblad jump operators. This is a stronger form of nonlinearity, supporting rich dynamical phenomena such as such Hopf bifurcations and strange attractors usually associated with classical nonlinear systems \cite{190709349}. State-dependent CPTP channels also support Bloch-ball {\it torsion}. Torsion can be created from the product of an SO(3) rotation generator $J_\mu$  with the projection of the Bloch vector along the twist axis. Abrams and Lloyd \cite{PhysRevLett.81.3992} and Childs and Young \cite{150706334} investigated state discrimination with $z$-axis torsion. K{\l}obus {\it et al.}~\cite{211113477} observed Feigenbaum's universal period doubling in a mean field model simulating torsion.
Torsion also arises in a qubit friendly extension \cite{211209005} of a rigorous duality between nonlinear mean field theory and the BBGKY hierarchy for $n$ interacting bosons in the $n \rightarrow \infty$ limit \cite{NachtergaeleJSP06,FrohlichCMP07,RodnianskiCMP09,ErdosJSP09,200605486}. 
Many of these nonlinear models  
come from mean field theory.

\clearpage

\begin{figure}
\includegraphics[width=12.0cm]{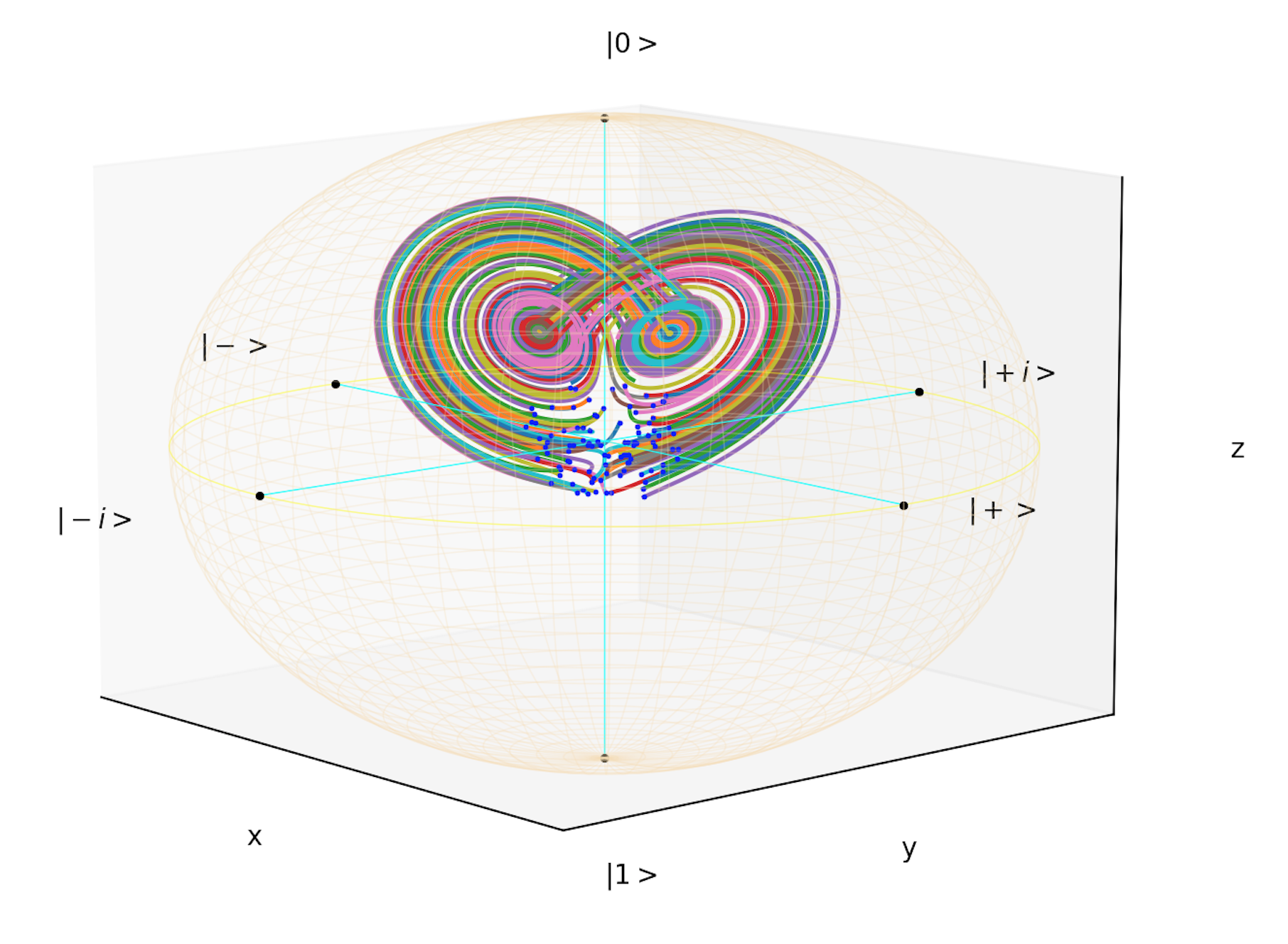} 
\caption{Bloch ball dynamics of the {\it Lor63} qubit. The faint yellow wireframe shows the Bloch sphere. Equator states $\KET{\pm} = 2^{-1/2} (\KET{0} \pm \KET{1} )$ and $\KET{\pm i} = 2^{-1/2} (\KET{0} \pm i \KET{1} )$ are also indicated with black dots and cyan lines. The model parameters used in the simulation are given in Table \ref{lorenz parameters table}.}
\label{lor63 figure}
\end{figure}

In this paper we investigate qubit PTP channels with both nonlinear $\phi$ and ${\rm tr}[\phi(X)] \neq 1$ that support generalized Lorenz attractors. The first version, which we call {\it Lor63}, implements Lorenz's 1963 model  \cite{LorenzJAS63}
\begin{eqnarray}
\frac{dx}{dt} &=& \sigma (y-x) , \\
\frac{dy}{dt} &=& \rho x - y - g x z,  \\
\frac{dz}{dt} &=& - \beta z + g x y ,
\end{eqnarray}
where ${\bf r} = (x,y,z) = {\rm tr}(X {\bm \sigma})$ is the Bloch vector. However here we increase the nonlinearity by a factor of $g \gg 1$ to shrink the attractor sufficiently as to contain it within the Bloch sphere. The master equation for the {\it Lor63} qubit in the Pauli basis is
\begin{eqnarray}
\frac{dX}{dt} = \frac{\sigma^a }{2}  
\bigg( \frac{dr^a}{dt} \bigg), \ \ 
\frac{dr^a}{dt}  = {\rm tr} \bigg( \! \frac{dX}{dt} \sigma^a \!\bigg)  \!  = G^{a b}({\bf r}) \, r^b  = ( L + g x J_x)^{a b} r^b \! ,
\label{lor63 qubit}
\end{eqnarray}
where $a, b \in \{ 1,2,3 \}$ and
\begin{eqnarray}
L = 
\begin{pmatrix}
-\sigma & \sigma & 0 \\
\rho & -1 & 0 \\
0 & 0 & -\beta \\
\end{pmatrix} \!
= L_{+} + L_{-}, \ \ 
L_{+} \! = \! 
\begin{pmatrix} 
-\sigma &  \frac{\rho + \sigma}{2} & 0 \\ 
 \frac{\rho + \sigma}{2}  & -1 & 0  \\ 
0 & 0 & -\beta \\ 
\end{pmatrix} \!
=  \big(  { \frac{\rho + \sigma}{2}} \big) \,  \lambda_1 - D, 
\end{eqnarray}
\begin{eqnarray}
L_{-}   \! = \!  \big(  \frac{\rho - \sigma}{2} \big)  \, J_z , \ \ 
\lambda_1  \! = \!  
\begin{pmatrix}
0 & 1 & 0 \\
1 & 0 & 0 \\
0 & 0 & 0 \\
\end{pmatrix} \! , \ \ 
D  \! = \!  
\begin{pmatrix}
\sigma & 0 & 0 \\
0 & 1 & 0 \\
0 & 0 & \beta \\
\end{pmatrix} \! , \ \ 
J_{x}  \! = \!  
\begin{pmatrix}
0 & 0 & 0 \\
0 & 0 & -1 \\
0 & 1 & 0 \\
\end{pmatrix} \! ,
\ \ 
J_{z}  \! = \!  
\begin{pmatrix}
0 & -1 & 0 \\
1 & 0 & 0 \\
0 & 0 & 0 \\
\end{pmatrix} \! . \ \ \ \ \ 
\end{eqnarray}
$X \in {\mathbb C^{2 \times 2}}$ is a Hermitian positive-semidefinite matrix with unit trace. Model parameters $\rho$, $\sigma$, $\beta$, $g$ are given in Table \ref{lorenz parameters table}. The nonlinear generator $G^{ab}({\bf r})$ is a $3  \times 3$ real matrix that depends on the Bloch vector ${\bf r}$. We decompose it into a linear (${\bf r}$ independent) operator $L$ plus $x$-axis torsion. The $J^{\rm ' s}$ are SO(3) generators: $ (J_a)_{bc} =  - \varepsilon_{abc}$ with $\varepsilon$ the Levi-Civita symbol. $L$ is decomposed into symmetric and antisymmetric parts implementing a non-Hermitian Hamiltonian $i L$. $\lambda_1$ is a Gell-Mann matrix. Note that $\lambda_1$ has a positive eigenvalue corresponding to an amplifying and entropy {\it decreasing} non-CP process \cite{KowalskAP19}. Techniques for constructing Gell-Mann matrices and other symmetric generators from jump operators are given in \cite{211105977}.
The {\it Lor63} qubit is simulated in Fig.~\ref{lor63 figure}. The blue points indicate random initial conditions. Trajectories rapidly approach one of the two disc-shaped sets (pink or cyan) and bounce back and forth between them in an unpredictable manner, mirroring the aperiodic reversals of the Malkus waterwheel lying in its Fourier representation \cite{Sparrow1982}.

\begin{table}[htb]
\centering
\caption{{\it Lor63} model parameters.}
\begin{tabular}{|c|c|c|}
\hline
  & Original & Here \\
 \hline
 $\rho$ &  28 & 28\\
 \hline
 $\sigma $ & 10 & 10 \\
 \hline
 $\beta $ & 8/3 & 8/3 \\
 \hline
 $g$ & 1 & 80 \\
 \hline
\end{tabular}
\label{lorenz parameters table}
\end{table}

\clearpage

\begin{figure}
\includegraphics[width=12.0cm]{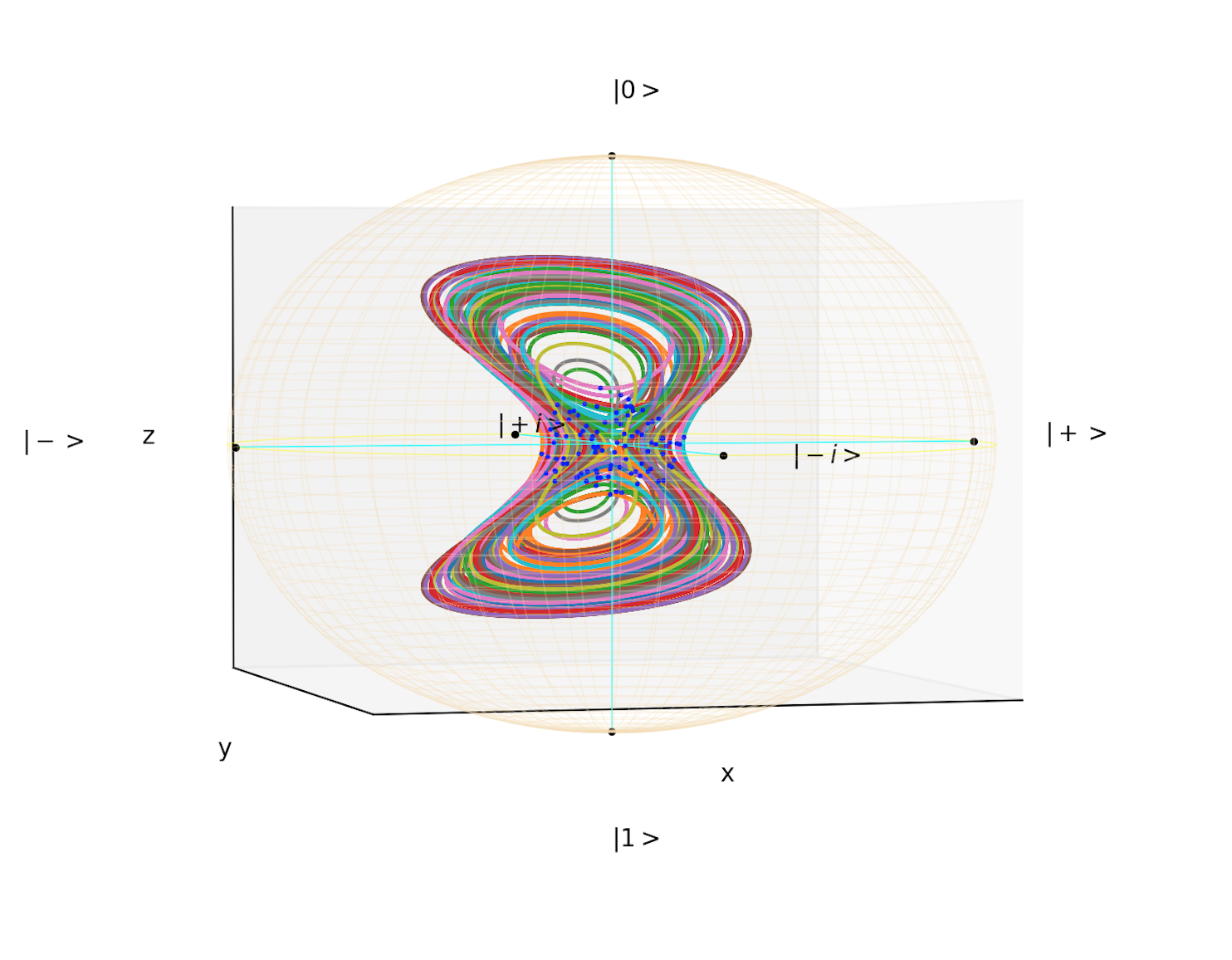} 
\caption{Bloch ball dynamics of the {\it GP butterfly} qubit. Blue dots indicate random initial conditions.}
\label{gp butterfly figure}
\end{figure} 

A Lorenz-like attractor can also be created from the $z$-axis torsion coming from the Gross-Pitaevskii (GP) equation \cite{13030371,13107301,150706334}, leading to an aesthetic attracting set shown in Fig.~\ref{gp butterfly figure}. We call this channel the {\it GP butterfly}. The {\it GP butterfly} qubit has an especially simple master equation:
\begin{eqnarray}
\frac{dr^a}{dt}  = {\rm tr} \bigg( \! \frac{dX}{dt} \sigma^a \!\bigg)  \!  = G^{a b}({\bf r}) \, r^b  = ( m \lambda_4 + g z J_z)^{a b} r^b \! ,
\label{gp butterfly}
\end{eqnarray}
where $m=10$, $g=40$, and 
\begin{eqnarray}
\lambda_4  =   
\begin{pmatrix}
0 & 0 & 1 \\
0 & 0 & 0 \\
1 & 0 & 0 \\
\end{pmatrix}
\end{eqnarray}
is another Gell-Mann matrix. The symmetric generator $\lambda_4$ can be implemented with Lindblad jump operators \cite{211105977}. 

\clearpage

In conclusion, we have proposed nonlinear PTP channels for the generation of Lorenz-like attractors in the Bloch ball. Despite its early prominence the Lorenz system defied rigorous analysis until rather recently when, in 2002, Tucker \cite{TuckerFCM02} established the existence of a strange attractor. Classical electrical circuits have been used to implement the Lorenz attractor and other chaotic and hyperchaotic attractors \cite{CuomoPRL93,QiCSF09,LiuEntropy19,TianComplexity21}, which might find cryptographic application \cite{200611847,12013114}. It is tempting to speculate that chaotic attractors will find application in quantum technology as well. However it is important to recognize the very large nonlinear coupling strengths required, making experimental realization especially challenging.

\acknowledgements

This work was partly supported by the NSF under grant no.~DGE-2152159.

\bibliography{Paper.bbl}

\end{document}